\documentstyle[12pt,latexsym,amsmath,graphicx,psfrag,fancyhdr]{article}

\begin{document}

\begin{titlepage}
\vspace*{1cm}
\begin{center}
{\Large \bf Hawking Radiation from a $(4+n)$-Dimensional\\[1mm] Rotating Black Hole on the Brane}

\bigskip \bigskip \medskip
{\large C.M. Harris$^{1}$ and P. Kanti$^2$}\\[0.5cm]

$^{1}${\it Cavendish Laboratory, University of Cambridge, Madingley Road,
Cambridge CB3 0HE, United Kingdom}\\[.4cm]
$^{2}${\it Department of Mathematical Sciences, University of Durham,
Science Site, South Road, Durham DH1 3LE, United Kingdom}
\\[1cm]

{\bf Abstract} \end{center}
\noindent
We study the emission of Hawking radiation in the form of scalar fields from a
$(4+n)$-dimensional, rotating black hole on the brane. We perform a numerical analysis
to solve both the radial and angular parts of the scalar field equation, and derive
exact results for the Hawking radiation energy emission rate. We find that,
in 5 dimensions, as the angular momentum increases, the emission rate is suppressed
in the low-energy regime but significantly enhanced in the intermediate and high-energy
regimes. For higher values of $n$, the Hawking radiation emission rate on the brane is
significantly enhanced, with the angular momentum, over the whole energy regime.
We also investigate
the energy amplification due to the effect of super-radiance and demonstrate that, in
the presence of extra dimensions, this effect is again significantly enhanced.

\end{titlepage}

During the past few years, great interest has been drawn to the concept of
trans-planckian collisions which can be realized in the framework of theories with
Large Extra Dimensions \cite{ADD}, under the assumption that the colliding particles 
have a center-of-mass energy $s$ larger than the $(4+n)$-dimensional, fundamental Planck
scale $M_*$. During such high-energy collisions, it is natural to expect that the products
will no longer be ordinary particles but heavy objects, arising in the context of
a fundamental theory of interactions including gravity \cite{Banks}.

Small, $(4+n)$-dimensional black holes are one of the possible products of such collisions
\cite{Banks, GT, DL}. As long as their mass $M_\text{BH}$ is larger than a few times the
fundamental Planck scale $M_*$, these black holes can be treated as classical, with
all the laws of black hole physics applying to them. One of the most important characteristics
would then be their decay in time, through the emission of Hawking radiation \cite{Hawking},
a characteristic that will also be the most striking observable effect of their existence.
The produced black holes, after shedding any additional quantum numbers inherited
from the colliding particles, will settle down to Kerr-like, rotating black holes, and
a {\it spin-down phase} will commence \cite{GT}, in which the black hole angular momentum 
will be gradually lost through the emission of Hawking radiation and super-radiance. A
{\it Schwarzschild phase} describing a non-rotating black hole will follow next 
with the emission of Hawking radiation resulting in the decrease of the black hole mass. 

The emission of Hawking radiation during the Schwarzschild phase has been studied both
analytically \cite{KMR, FS1} and numerically \cite{HK} (see \cite{Kanti}, for a review
and related works), however the same is not true for the spin-down phase which has been,
apart from a few exceptions \cite{FS2,IOP}, largely ignored. In this short
article, we present, for the first time in the
literature, exact numerical results for the Hawking radiation energy emission rate 
coming from a $(4+n)$-dimensional, rotating black hole. We focus our attention on the
emission of scalar field radiation on the brane where, according to the assumptions of
the model \cite{ADD}, the observer is situated. After formulating the problem, we
numerically solve both the angular and radial scalar field equations to determine the
exact angular eigenvalues and radial wavefunction, respectively, for arbitrarily large
black hole angular momentum and energy of the emitted particles. Then, we derive
the absorption coefficient and subsequently the differential energy emission rate for
Hawking radiation and the energy spectrum amplification due to the super-radiance effect.


The line-element describing a $(4+n)$-dimensional, rotating, uncharged black hole
was found by Myers and Perry \cite{MP}. A black hole created by the collision of
particles moving in a $(4+n)$-dimensional spacetime can have up to $(n+3)/2$ angular
momentum parameters. However, in the context of the theory with Large Extra Dimensions
\cite{ADD}, the colliding partons are restricted to propagate on an infinitely-thin
3-brane and therefore they
have a non-zero impact parameter only on a 2-dimensional plane along our brane; thus,
it is reasonable to assume that they will acquire only one non-zero angular parameter
about an axis in the brane. The 4-dimensional induced spacetime on the brane, in
which the emitted particles propagate, takes the form \cite{Kanti}
\begin{equation}
\begin{split}
ds^2=\left(1-\frac{\mu}{\Sigma\,r^{n-1}}\right)dt^2&+\frac{2 a\mu\sin^2\theta}
{\Sigma\,r^{n-1}}\,dt\,d\varphi-\frac{\Sigma}{\Delta}dr^2 \\[3mm] &\hspace*{-3cm}
-\Sigma\,d\theta^2-\left(r^2+a^2+\frac{a^2\mu\sin^2\theta}{\Sigma\,r^{n-1}}\right)
\sin^2\theta\,d\varphi^2, 
\end{split} \label{induced}
\end{equation}
where
\begin{equation}
\Delta=r^2+a^2-\frac{\mu}{r^{n-1}} \quad\mbox{ and } \quad\Sigma=r^2+a^2\cos^2\theta\,.
\label{master}
\end{equation}
The parameters $\mu$ and $a$ are related to the mass and angular momentum, respectively,
of the black hole through the definitions \cite{MP}
\begin{equation}
M_\text{BH}=\frac{(n+2) A_{n+2}}{16 \pi G}\,\mu\,, \qquad
J=\frac{2}{n+2}\,M_\text{BH}\,a\,. \label{parameters}
\end{equation}
In the above, $G$ is the $(4+n)$-dimensional Newton's constant, and $A_{n+2}$ the area
of a $(n+2)$-dimensional unit sphere given by: $A_{n+2}=2 \pi^{(n+3)/2}/\Gamma[(n+3)/2]$.
Note that the induced line-element on the brane has an explicit dependence on the number
$n$ of extra dimensions.

The black hole horizon is given by solving $\Delta(r)=0$.  Unlike the 4D Kerr black hole
for which there is an inner and outer solution for $r_\text{h}$, for $n\geq1$ there is only
one solution of this equation. In addition, in the $n=0$ and $n=1$ cases, there is a maximum
possible value of $a$, otherwise there are no solutions of $\Delta=0$ and thus no horizon
to shield the singularity at $r=0$. On the other hand, for $n>1$ there is no fundamental
upper bound on $a$ and a horizon $r_\text{h}$ always exists. For general $n$, the horizon
radius is given by $r_\text{h}^{n+1}=\mu/(1+a_*^2)$, where we have defined
$a_*=a/r_\text{h}$. An upper bound can nevertheless be imposed on the angular momentum
parameter of the black hole by demanding the creation of the black hole itself from
the collision of the two particles. The maximum value of the impact parameter between
the two particles that can lead to the creation of a black hole was found to be
\cite{harris}
\begin{equation}
b_{\rm max} =2\,\left[1+\left(\frac{n+2}{2}\right)^2\right]^{-\frac{1}{n+1}}\,
\mu^{\frac{1}{n+1}}\,. \label{bmax}
\end{equation}
If we, then, write $J=b M_{\rm BH}/2$ \cite{IOP} for the angular momentum of the black hole,
and use the second of Eq. (\ref{parameters}), we obtain \cite{harris}
\begin{equation}
a^{\rm max}_*=\frac{n+2}{2}\,.
\label{amax}
\end{equation}

By using the Newman-Penrose formalism, the equation for the propagation of a field with
spin 0, 1/2 and 1, in the gravitational background induced on the brane, can be found
\cite{Kanti}.
Here, we will focus on the emission of scalar fields leaving the analysis for non-zero spin
fields for a subsequent work \cite{HKW}. By using the field factorization 
\begin{equation}
\phi(t,r,\theta,\varphi)= e^{-i\omega t}\,e^{i m \varphi}\,R(r)\,T^{m}_{\ell}(\theta)\,,
\end{equation}
where $T^{m}_{\ell}(\theta)$ are the so-called spheroidal harmonics \cite{Flammer}, we
obtain a set of decoupled radial and angular equations,
\begin{equation}
\frac{d}{dr}\biggl(\Delta\,\frac{d R}{dr}\biggr)+\left(\frac{K^2}{\Delta} -\Lambda^m_{\ell}
\right)R=0\,, \label{radial}
\end{equation}
\begin{equation}
\label{spinang}
\frac{1}{\sin\theta} \frac{d}{d\theta}\left(\sin\theta\,\frac{d T^m_{\ell}(\theta)}{d\theta}
\right) + \biggl(-\frac{m^2}{\sin^2\theta} 
+a^2\omega^2\cos^2\theta + E^m_{\ell}\biggr) T^m_{\ell}(\theta)=0\,,
\end{equation} 
respectively, where we have defined
\begin{equation}
K=(r^2+a^2)\,\omega-am\,, \qquad
\Lambda^m_{\ell}=E^m_{\ell}+a^2\omega^2-2am\omega\,.
\end{equation}

The Hawking temperature of the $(4+n)$-dimensional, rotating black hole is found to be
\begin{equation}
T_\text{H}=\frac{(n+1)+(n-1)a_*^2}{4\pi(1+a_*^2)r_\text{h}}\,,
\end{equation}
and leads to the emission of scalar Hawking radiation on the brane,
with the corresponding differential energy emission rate given by the expression 
\begin{equation}
\frac{d E(\omega)}{dt} = \sum_{\ell,m} |{\cal A}_{\ell,m}|^2\,
\frac{\omega}{\exp\left[(\omega-m\Omega)/T_\text{H}\right] - 1}\,\frac{d\omega}{2\pi}\,.
\label{rpower}
\end{equation}
In the above, the rotation velocity $\Omega$ is defined as
\begin{equation}
\Omega=\frac{a_*}{(1+a_*^2)\,r_\text{h}}\,,
\end{equation}
while $|{\cal A}_{\ell,m}|^2$ is the absorption probability for a scalar particle propagating
in the background induced on the brane (\ref{induced}). Its presence in the expression for the
emission rate modifies the blackbody profile of the spectrum due to its explicit dependence on
the energy $\omega$ of the emitted particle, its angular momentum numbers $(\ell, m)$, and the number
of extra dimensions $n$. The exact form of $|{\cal A}_{\ell,m}|^2$ can be found by solving the
radial equation (\ref{radial}), a task which we have performed by using numerical analysis. The
numerical solution obtained for $R(r)$ interpolates between the asymptotic solutions at the
horizon of the black hole and infinity. Near the horizon, Eq. (\ref{radial}) leads to the
asymptotic solution 
\begin{equation}
R_\text{h}(r^*)=A_1\,e^{i k r^*} + A_2\,e^{-i k r^*}\,,
\label{asy-hor}
\end{equation}
where $A_{1,2}$ are integration constants,
\begin{equation}
k=\omega-\frac{ma}{r_\text{h}^2+a^2}\,,
\end{equation}
and $r^*$ is the tortoise coordinate defined by
\begin{equation}
\frac{dr^*}{dr}=\frac{r^2+a^2}{\Delta(r)}\,.
\end{equation}
A boundary condition must be applied in the near-horizon regime, namely that the solution
must contain only incoming modes; this is satisfied if we set $A_1=0$. On the other hand,
for fixed $a_*$ and large $r$, the solution at infinity takes the form
\begin{equation}
R_\infty(r)=B_1\,\frac{e^{i \omega r}}{r} + B_2\,\frac{e^{-i \omega r}}{r}\,,
\label{asy-inf}
\end{equation}
where $B_{1,2}$ are again integration constants. 

\begin{figure}[t]
\begin{center}
\psfrag{y}[][][1.0]{$E^m_{\ell}(a\omega)$} \psfrag{x}[][][1.0]{$a\omega$}
\includegraphics[height=6.5cm]{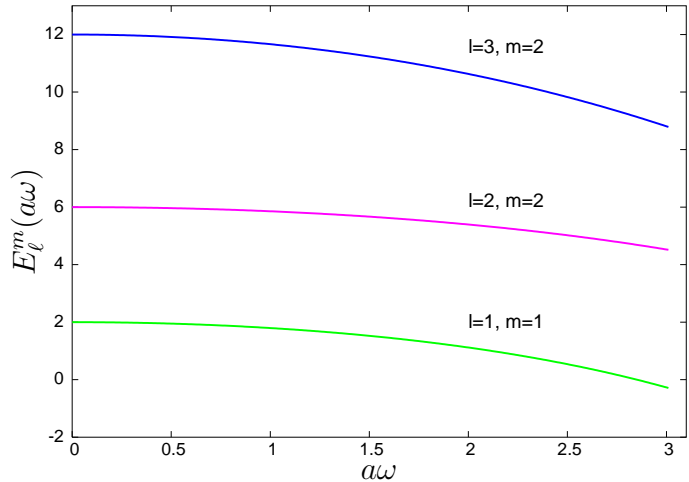}
\caption{{The angular eigenvalues $E^m_\ell$, for $s=0$, as a function of $a \omega$ for
three angular momentum modes: $\ell=m=1$, $\ell=m=~2$, and $\ell=3$, $m=2$. 
\hspace*{3.7cm}\label{eigen}}}
\end{center}
\end{figure}

However, before the numerical integration of Eq. (\ref{radial}) can take place, the value of
the constant $\Lambda^m_\ell$, or equivalently, the angular eigenvalue $E^m_{\ell}$, must be
determined. The eigenvalues of the spheroidal harmonics are functions of $a\omega$, and
an analytic form can be found only in the limit of small $a \omega$  \cite{Seidel} --
the only work in the literature on the Hawking radiation coming from a $(4+n)$-dimensional,
rotating black hole \cite{IOP} employs this approximate expression for $E^m_\ell$, and the
corresponding analysis is valid only in the limit of low energy and low black-hole angular
momentum. The exact value of $E^m_\ell$, for arbitrarily large values of $a \omega$ can be
obtained by using a continuation method \cite{Wasserstrom, harris}.
This is a generalization of perturbation theory which can be applied for arbitrarily large
changes in the initial Hamiltonian for which the eigenvalues are known. Here, we
briefly describe this technique for the case $s=0$. Equation (\ref{spinang})
can be alternatively written as
\begin{equation}
\label{opang}
({\mathcal H}_0+{\mathcal H}_1)\, T^m_{\ell}(\theta,a\omega) =-E^m_{\ell} (a\omega)\,
T^m_{\ell}(\theta,a\omega)\,, 
\end{equation}
where ${\mathcal H}_0$ stands for the first two terms of the differential operator in
Eq. (\ref{spinang}), and ${\mathcal H}_1=a^2\omega^2\cos^2\theta$. For $a \omega=~0$,
${\mathcal H}_1$ vanishes, and $T^m_{\ell}(\theta,0)$ reduce to the usual spherical
harmonics $S^m_{\ell}(\theta)$, with $E^m_{\ell}(0)=\ell(\ell+1)$. To employ the
continuation method, we write the $T^m_{\ell}(\theta,a\omega)$ functions in the basis
of the $\theta$-parts of the spherical harmonics:
\begin{equation}
T^m_{\ell}(\theta,a\omega) = \sum_{\ell'}\,B_{\ell\ell'}^m(a\omega)\,S_{\ell'}^m(\theta)\,.
\end{equation}
By differentiating Eq.~(\ref{opang}) and applying the same techniques as in perturbation
theory, we obtain the equation
\begin{equation}
\label{conte}
\frac{d \,E_{\ell}^m}{d(a\omega)}=-\sum_{\alpha,\beta}\,B_{\ell\alpha}^m \,
B_{\ell\beta}^m\,\langle\alpha|\beta\rangle\,,
\end{equation}
where $\langle\alpha|\beta\rangle \equiv \langle \alpha m|d{\mathcal H}_1/d(a\omega)|\beta m\rangle$.
The coefficients $B^m_{\ell a}$ satisfy themselves a similar differential equation, i.e.
\begin{equation}
\label{contb}
\frac{d \,B_{\ell\ell'}}{d(a\omega)}=-\sum_{\alpha,\beta,\gamma \neq l}\,
\frac{B_{\gamma\alpha} \, B_{\ell\beta}}{E^m_{\ell}-E_{\gamma}^m}\,
\langle\alpha|\beta\rangle\,B_{\gamma \ell'}\,,
\end{equation}
with initial condition  $B_{\ell\ell'}^m(0)=\delta_{\ell\ell'}$. By integrating
Eqs.~(\ref{conte}) and (\ref{contb}), it is possible to obtain the eigenvalues of the
spheroidal functions for any $\ell$ and $m$ and for arbitrarily large values of
$a\omega$.  This integration was performed numerically by using a Runge-Kutta method.
Figure~\ref{eigen} shows the angular eigenvalues $E_{\ell}^m$ for scalar
fields as a function of $a \omega$, for some indicative angular momentum modes.

Having derived the eigenvalues  $E_\ell^m$, we can now proceed to integrate Eq. (\ref{radial})
and derive the solution for the radial function $R(r)$. The integration starts at the horizon
of the black hole and proceeds towards infinity. Comparing our numerical results with the
asymptotic solution at infinity (\ref{asy-inf}), we determine the integration
constants $B_1$ and $B_2$.
The absorption
probability for scalar fields can then be derived from the relation
\begin{equation}
|{\cal A}_{\ell,m}|^2=1-|{\cal R}_{\ell,m}|^2=1-\biggl|\frac{B_1}{B_2}\biggr|^2\,,
\label{abs}
\end{equation}
where ${\cal R}_{\ell,m}$ is the reflection coefficient given by the ratio of the
outgoing and ingoing amplitudes at infinity.
\begin{figure}
\begin{center}
\psfrag{x}[][][1.0]{$\omega\,r_\text{h}$}
\psfrag{y}[][][1.0]{$r_\text{h}\,d^2E/dt\,d\omega$}
\includegraphics[height=6.5cm,clip]{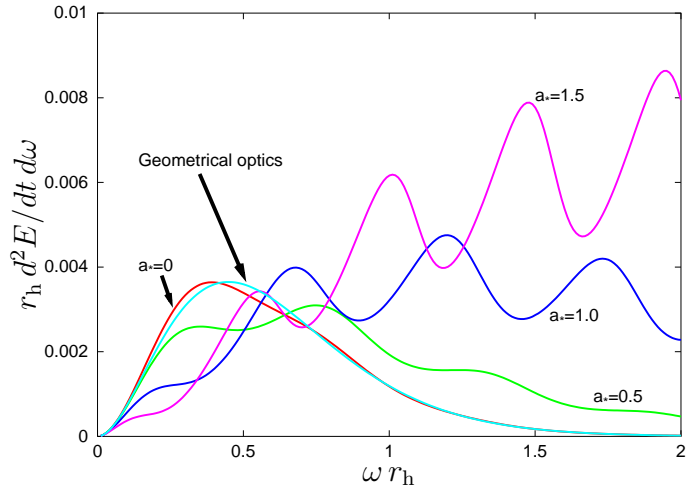}
\caption{Power spectra for scalar emission on the brane from rotating black holes,
for $n=1$ and various values of $a_*$.\hspace*{1.5cm}}
\label{s0a}
\end{center}
\end{figure}

The value of the absorption probability $|{\cal A}_{\ell,m}|^2$ is then inserted into
Eq. (\ref{rpower}) to determine the differential energy emission rate per unit time
and frequency by the black hole on the brane. This rate is shown in Fig.~\ref{s0a},
for the case of a 5-dimensional black hole ($n=1$). For convenience, we assume that
the black hole horizon value remains fixed, and set $r_\text{h}=1$. Figure \ref{s0a}
shows the emission spectrum on the brane for various values of
the angular momentum parameter $a_*$, up to the value $a_*^{\rm max}=1.5$ defined
by the black-hole-creation constraint\footnote{In terms
of fixed mass parameter $\mu$, the maximum value of $a$ is
$0.83\,\sqrt{\mu}$. Had the angular momentum parameter been allowed to
increase indefinitely, the critical value for the existence of the horizon, i.e.
$a=\sqrt{\mu}$, would have been reached. For this value, both the black hole horizon
and the temperature vanish, the latter leading to the suspension of the emission of
Hawking radiation. The non-vanishing value of the area of the black hole induced on
the brane, given by $A_\text{H}^{(4)}=4 \pi (r_\text{h}^2+a^2)$,
as opposed to the vanishing one of the $(4+n)$-dimensional black hole, given by
$A_\text{H}^{(4+n)}=\Omega_{2+n}\,r_\text{h}^n\,(r_\text{h}^2+a^2)$ [with
$\Omega_{2+n}$ the solid angle of the $(2+n)$-dimensional space], would lead to a
substantial suppression of the emission of energy in the bulk compared to
the one on the brane during these last stages, in addition to the suppression
found in the non-rotating case \cite{HK, EHM}.}
(\ref{amax}). The dimensionless parameter
$\omega r_\text{h}$ on the horizontal axis ade\-quately
covers the low, intermediate and high-energy regimes. The different curves in the figure
allow us to compare the energy emission rates for black holes with the same horizon radius
and different angular momentum. When $r_\text{h}$ is kept fixed, as in this case, the
temperature of the black hole decreases sharply with $a_*$, leading to the observed
suppression of the energy emission rate in the low-energy regime -- this is in agreement
with the behaviour found by analytical methods in the low $\omega$ and low $a$ limit
\cite{IOP}. As the energy increases further, however, the absorption probability is
significantly enhanced; this enhancement gradually overcomes the decrease in the black
hole temperature leading to the observed increase in the energy emission rate, as
$a_*$ increases, both in the intermediate and high-energy regimes. This in turn leads
to a significant enhancement of the total emissivity of a 5-dimensional rotating
black hole (that is, energy emitted per unit time over the whole frequency band) compared
to that of a non-rotating black hole of the same dimensionality. 

The numerical results produced above allowed the statement made in \cite{GT}, 
according to which the emission of Hawking radiation is dominated by modes with $\ell=m$,
to be tested.  We have found that Fig.~\ref{s0a} looks the same at the $\sim$~90\% level if
only the $\ell=m$ modes are included in the sum of Eq.~(\ref{rpower}), a result that confirms
this statement.

Keeping the black hole horizon value $r_\text{h}$ fixed during our analysis was a
convenient choice from a calculational point of view. However, as $a$ varies, this
leads to the comparison of energy emission rates for black holes with different masses.
From a phenomenological point of view, fixing the mass parameter $\mu$ of the black hole,
instead, makes more sense. In Fig. 3, we present the energy emission spectrum on the
brane for a 6-dimensional black hole, i.e. for $n=2$. The angular momentum parameter
now varies from zero to the maximum value -- derived from Eqs. (\ref{bmax})-(\ref{amax}) --
of $a^{\rm max}=1.17\,M_*^{-1}$, where $M_*$ is the fundamental
Planck scale. For fixed $\mu$, the black hole temperature again decreases\footnote{To
be exact, the temperature decreases up to the point $a=1.1\,M_*^{-1}$, and then
starts increasing up to $a^\text{max}=1.17\,M_*^{-1}$; however, the increase in
its value is only 0.5\%, which leads to no observable effect.}, as $a$ increases;
however, the increase in this case is only a mild one, which allows for the enhancement
of the absorption probability to dominate the spectrum. This leads to a significant 
enhancement of the energy emission rate on the brane in all energy regimes, and thus
to a clear enhancement of the total emissivity of a rotating black hole compared to the one
of a non-rotating black hole with the same mass. 

\begin{figure}[t]
\begin{center}
\psfrag{x}[][][1.0]{\hspace*{3.5cm}$\omega\,\,\,\,\,\,({\rm in\,\,units\,\,of}\,\,M_*)$}
\psfrag{y}[][][1.0]{$r_\text{h}\,d^2E/dt\,d\omega$}
\includegraphics[height=6.5cm,clip]{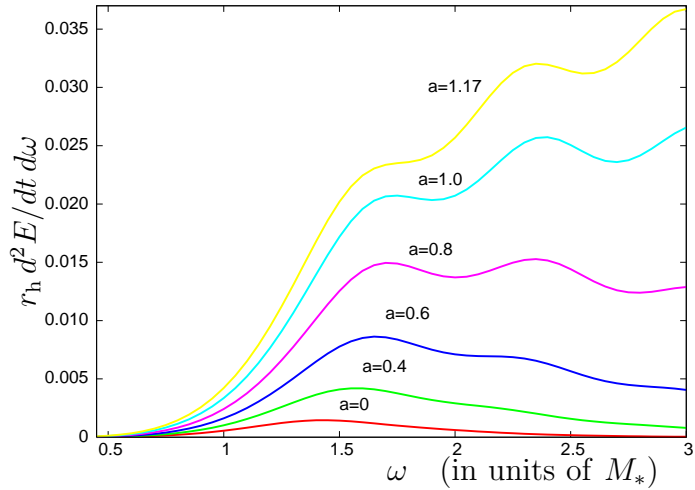}
\caption{Power spectra for scalar emission on the brane from rotating black holes,
for $n=2$ and various values of $a$ in units of $M_*^{-1}$.\hspace*{1.5cm}}
\label{s0n2}
\end{center}
\end{figure}
\begin{figure}[t]
\begin{center}
\mbox{\psfrag{x}[][][1.0]{$\omega/m\Omega$}
\psfrag{a}[][][0.8][18]{\textsf{l=m=1}}
\psfrag{b}[][][0.8][30]{\textsf{l=m=2}}
\psfrag{c}[][][0.8][37]{\textsf{l=m=3}}
\psfrag{d}[][][0.8][40]{\textsf{l=m=4}}
\psfrag{e}[][][0.8][45]{\textsf{l=m=5}}
\psfrag{f}[][][0.8]{\textsf{l=2, m=1}}
\includegraphics[height=7.3cm]{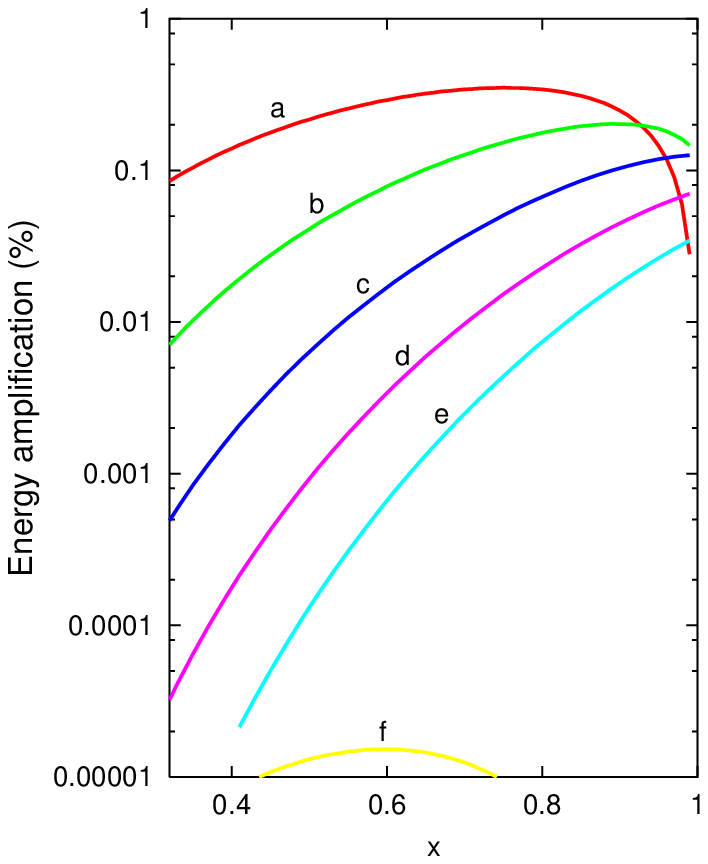}} \hspace*{0.5cm}
{\psfrag{x}[][][1.0]{$\omega/m\Omega$}
\psfrag{a}[][][0.8][18]{\textsf{l=m=1}}
\psfrag{b}[][][0.8][28]{\textsf{l=m=2}}
\psfrag{c}[][][0.8][35]{\textsf{l=m=3}}
\psfrag{d}[][][0.8][40]{\textsf{l=m=4}}
\psfrag{e}[][][0.8][43]{\textsf{l=m=5}}
\psfrag{f}[][][0.8]{\textsf{l=2, m=1}}
\includegraphics[height=7.3cm]{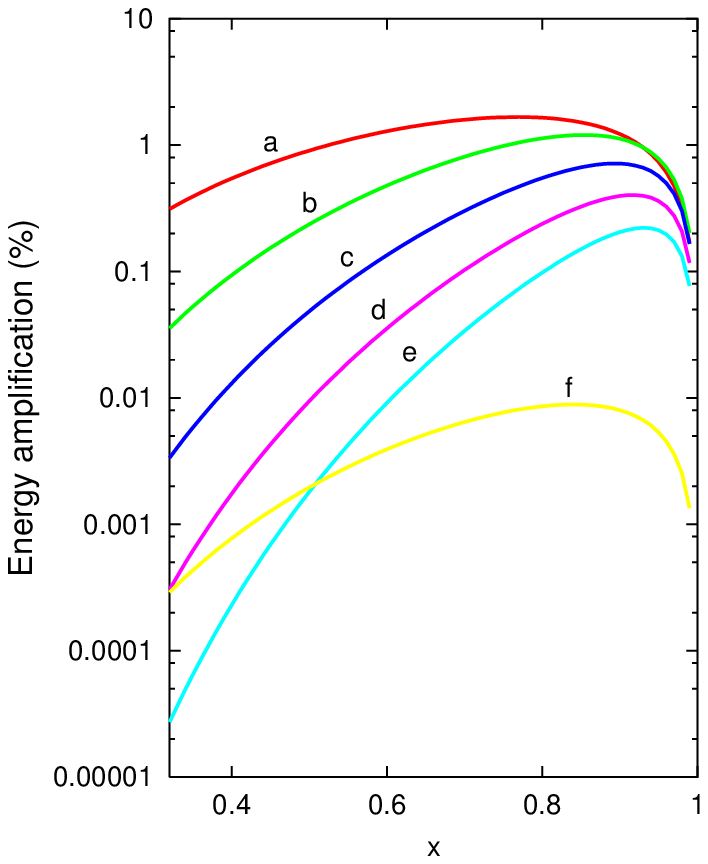}}
\caption{{\bf (a)} Super-radiant scattering of scalars by a maximally rotating
($a_*=1$) 4-dimensional black hole; {\bf (b)} Super-radiant scattering of scalars
by an induced-on-the-brane 6-dimensional black hole with $a_*=1$.
\hspace*{2.0cm}\label{super4D-6D}}
\end{center}
\end{figure}
An interesting effect which takes place during the propagation of a bosonic field in the
background of a rotating black hole is super-radiance \cite{super}, that is, the amplification
of the
amplitude of the incident wave. This becomes manifest when the reflection probability becomes
larger than unity, or equivalently when the absorption probability becomes negative. In the
context of a higher-dimensional model, the only studies in the literature are an analytic
approach \cite{FS2} which confirmed super-radiance for bulk scalars incident on five-dimensional
black holes, and a sole numerical result for $n=6$ and $\ell=m=1$ \cite{IOP2}. 
In the context of our numerical analysis, we have investigated this effect for various
values of the angular momentum of the black hole and number of extra dimensions.
In Fig. \ref{super4D-6D}, we compare the energy amplification due to
super-radiant scattering of scalar fields for a 4-dimensional, maximally-rotating black
hole with $a_*=1$ (equivalent to $a=M_\text{BH}$), and a 6-dimensional black hole with
again $a_*=1$. The vertical axis is the
absorption probability (in fact $-|\hat {\cal A}_{\ell,m}|^2$) expressed as a percentage
so that it gives the percentage energy amplification of the incident wave.
Figure~\ref{super4D-6D}(a) shows excellent agreement with the results produced for $n=0$ and
$a_*=1$ in \cite{Press:1972}. Comparing the vertical axes of Figs.~\ref{super4D-6D}(a) and
\ref{super4D-6D}(b), we clearly see that, in the
presence of extra dimensions, the peak amplification is more significant. For small
values of $a_*$, it is found that the $\ell=m=1$ mode provides the greatest amplification
as in the 4-dimensional case. However, for larger values of $a_*$ or $n$, the maximum
amplification can occur in modes with larger values of $\ell$: for $n=6$, and $a_*=4.0$,
the maximum energy amplification is found to be around 9\% and occurs in the
$\ell=m=7$ mode.

In summary, in this work we have studied the emission of Hawking radiation in the form
of scalar fields from a $(4+n)$-dimensional, rotating black hole on the brane. We have
performed a numerical analysis to solve both the radial and angular parts of the scalar
field equation on the brane, and derived exact results for the Hawking radiation energy
emission rate. We found that in the case of a 5-dimensional black hole, the emission
rate is suppressed in the low-energy regime, as the angular momentum increases -- in
agreement with previous approximate results -- but is significantly enhanced in the
intermediate and high-energy regimes, that were until now unexplored. We have extended
our analysis to black holes of higher-dimensionality and, as an illustrative example, we
have presented the spectrum of a 6-dimensional black hole: in this case, the energy
emission rate on the brane is enhanced with the angular momentum over the whole
energy band, a behaviour that persists for all higher values of $n$. We have
also investigated the amplification of the incident wave due to the effect of
superradiance, and showed that this effect is also significantly
enhanced in the presence of extra dimensions.


{\bf Acknowledgments}
We would like to thank B. Webber and E. Winstanley for useful discussions. During most
of this work, C.M.H. was funded by the UK PPARC Research Studentship PPA/S/S/2000/03001.
The work of P.K. was funded by the UK PPARC Research Grant PPA/A/S/2002/00350.

\end{document}